\documentclass[12pt]{iopart}

\usepackage{graphicx}

\begin{document}
\title[Nguyen \etal]{Energy dependent relative charge transfer cross sections of Cs$^{+}$+ Rb(5s, 5p)}

\author{H.~Nguyen}
\address{Department of Physics, University of Mary Washington, Fredericksburg, VA, 22401, USA}
\author{R.~Br\'{e}dy}
\address{Institut Lumi\`{e}re Mati\`{e}re, UMR5306 Universit$\acute{e}$ Lyon 1-CNRS, Universit$\acute{e}$ de Lyon, 69622 Villeurbanne cedex, France }

\author{X.~Fl\'{e}chard}
\address{ LPC Caen, ENSICAEN, Universit$\acute{e}$ de Caen Basse Normandie, CNRS/IN2P3-ENSI, Caen, France}
\author{B.~D.~DePaola}
\address{Department of Physics, Kansas State University, Manhattan, Kansas, 66506, USA}

\ead{hnguyen@umw.edu}

\begin{abstract}
Magneto optical trap recoil ion momentum spectroscopy is used to measure energy-dependent charge exchange cross sections in the Cs$^{+}$ + Rb(5s,5p) system over a range of projectile energies from 3.2 keV to 6.4 keV. The measurements are kinematically complete and yield cross sections that are differential in collision energy, scattering angle, and initial and final states.
\end{abstract}


\maketitle

\section{Introduction:}
Understanding charge exchange processes is of importance to a broad range of pure and applied research. For this reason charge transfer processes in collisions of single and multi-charged ions with ground and excited target atoms have been the subject of extensive theoretical and experimental studies. The important role of these processes in many laboratory and astrophysical plasmas stems from their large cross sections, which are proportional to the ionic charge, and pronounced final-state-selectivity. With increasing ionic charge, higher excited states are populated in the electron transfer process, the decay radiation from which can serve as a useful plasma diagnostic tool.  An example of this is the widely used charge-exchange-recombination spectroscopy diagnostic in magnetic fusion plasmas~\cite{Fonck1985}. The electron capture processes of multiply charged impurity ions with neutral hydrogen species are also important in the studies of impurity transport in the edge and diverter plasma regions~\cite{Isler1994}.

In the late 1960's Perel and coworkers, using then state-of-the-art
techniques, were able to detect distinct oscillations in the total
charge transfer cross sections as a function of collision energy in
alkali ion-alkali atom systems~\cite{Perel1965}. Target-atom excited states were attributed to polarization excitation by the incident ion during the initial phases of the collision. However, investigations of collisions of cesium projectiles with excited target rubidium at these energies were not possible due to the technical difficulties at the time to create a well-characterized excited state target.

Since the advent of the laser and laser cooling, excited states of rubidium targets can be routinely prepared for this type of collision. Studies of energy dependent state specific charge exchange processes are now possible using Magneto Optical Trap Recoil Ion Momentum Spectroscopy (MOTRIMS) \cite{Flechard2001,VPoel2001,Hoekstra2001,Nguyen2004,DePaola2008,Blieck2008}. This new technique, an outgrowth of the more general COLTRIMS (COLd Target Recoil Ion Momentum Spectroscopy)\cite{Ullrich1997,Dorner2000,Moshammer1996,Kollmus1997} method, enables high resolution Q-value and scattering angle measurements in ionizing ion-atom collisions.  Of paramount importance in this technique is that the thermal momentum distribution of the target be small compared to the momentum transferred to it in the collision. In the COLTRIMS method, this is typically accomplished by pre-cooling the target, and then allowing it to undergo supersonic expansion.  The technique is a powerful one and a tremendous amount of collisions physics has been understood through its use\cite{Ullrich1997,Dorner2000}. In the MOTRIMS technique, the supersonically expanded gas target is replaced by a magneto optical trap which yields several advantages. First, the atoms can be laser cooled to a far lower temperature than through supersonic expansion, thereby allowing improved momentum resolution, now limited only by detector time and position resolution.  Second, target atoms which are unsuitable for supersonic expansion, such as the alkalis and alkali earth elements, are readily laser cooled and trapped. Third, these target species are of necessity readily excited by lasers, allowing for collisions studies on excited as well as ground state targets.

Recently, using a combination of a Magneto Optical Trap and a Reaction Microscope~\cite{Fischer2012}, kinematically complete break-up processes of laser-cooled atoms were measured. The dynamics in swift ion-atom collision of 1.5~MeV/amu O$^{8+}$ + Li were studied at an unprecedented level of precision and detail. In a different experiment, using MOTRIMS, low-energy single charge transfer Na$^{+}$+$^{87}$Rb(5s, 5p) collisions were investigated using trapped Rb atoms as the target~\cite{Leredde2012}. High resolution, fully differential cross sections in scattering angle, initial state, and final state of the system were measured with an extraction of the recoil ions transverse to the ion beam axis and a fast switch for the MOT magnetic fields. Three dimensional recoil-ion momentum reconstruction provided accurate relative cross sections for the active channels and clearly resolved their associated distributions in projectile scattering angle. Experimental results were found to be in excellent agreement with molecular close-coupling calculations. Also, using similar MOTRIMS techniques, single electron transfer and ionization in collisions of N$^{5+}$ and Ne$^{8+}$ with Na(3s, 3p) were investigated both experimentally and theoretically at collision energies from 1 to 10~keV/amu~\cite{Blank2012,Otranto2012}. Relative cross section magnitudes and energy dependencies were found to be in good agreement with classical-trajectory Monte Carlo calculations.

In this work, MOTRIMS is used to measure charge exchange cross sections that are energy-dependent, kinematically complete, differential in initial state, final state and scattering angles, for the dominant channels over a range of projectile energies in the Cs$^{+}$ + $^{87}$Rb(5s, 5p) system. The data are suitable for use as a check of future theoretical calculations; for example using the two center atomic or molecular orbital close-coupling method~\cite{Leredde2012,Lee2002}.

\section{Experimental setup:}
Magneto optical trap-recoil ion momentum spectroscopy is the principal tool used in these experiments to measure the recoil ion's 3-dimensional momentum vector through time-of-flight (TOF) and 2-dimensional position-sensitive detection (2D-PSD). Q-values and scattering angles are determined from this measured momentum vector. Using the Kansas State University MOTRIMS apparatus, we performed our experiments using the longitudinal extraction method~\cite{Nguyen2004, NguyenThesis}. The relative advantages of transverse versus longitudinal experimental design for extractions is well discussed in the literature\cite{Ullrich1997,Dorner2000,Moshammer1996,Kollmus1997}. For the experiment described here the recoil ion TOF was approximately $70~\mu$sec, and was measured with a precision of about 2~nsec. The slight energy spread in the projectile beam, which was of the order of 1~eV out of a few~keV, determined the limits on resolution for the TOF measurement. The small angle (3.5~degrees) between the projectile beam axis and spectrometer extraction field was accounted for in the analysis; uncertainty in this angle had negligible effects on the timing or spatial precision.

Longitudinal and transverse momentum components were deduced through recoil TOF and position on the 2D-PSD.  The Q-value of the collision, defined as the difference between the system's total binding energy before and after the collision, is found through conservation of energy and momentum and is given by
\begin{equation}
\label{Eq:QV}
Q=-\frac{m_{e}v_{P_{initial}}^{2}}{2}- v_{P_{initial}}P_{R_{\parallel}} ,
\end{equation}
where non-subscripted $P$ indicates momentum, and the subscripts $P$, $R$, and $e$ refer to projectile, recoil, and electron, respectively.  Thus, as Eq.~[\ref{Eq:QV}] shows, one can express the collision Q-value in terms of the recoil ion longitudinal momentum.

The projectile scattering angle is given by
\begin{equation}
\label{Eq:consv_E_p_perp}
\tan \theta_P = \frac{P_{P_{\perp}}}{P_{P_{initial}}-P_{R_{\parallel}}} ~, \end{equation}
where $\theta_P$ is the polar projectile scattering angle. In Eqs.~[\ref{Eq:QV}] and [\ref{Eq:consv_E_p_perp}], the subscripts $\parallel$ and $\perp$ refer to the components of momentum that are parallel and perpendicular to the collision axis, respectively.

In general, the momentum equations must also account for the momentum of any ejected electrons. However in these measurements we are investigating the $Cs^+ + Rb \rightarrow Cs + Rb^+$ collision processes for which no electrons are emitted into the continuum. Therefore, in addition to Eqs.~[\ref{Eq:QV}] and~[\ref{Eq:consv_E_p_perp}], we have
\begin{equation}
\label{Eq:proj_recoil_mom}
-P_{R_{\perp}}=P_{P_{\perp}} .
\end{equation}

Because $P_{P_{initial}} \gg P_{R\parallel}$,
Eq.~[\ref{Eq:consv_E_p_perp}] becomes, in the small angle
approximation,
\begin{equation}
\label{perp} \theta_P=\frac{-P_{R_{\perp}}}{P_{P_{initial}}} ~,
\end{equation}
where we have used Eq.~[\ref{Eq:proj_recoil_mom}]. Notice that
Eq.~[\ref{perp}] shows that the projectile scattering
angle can be characterized by the transverse component of the recoil
ion momentum while the component of recoil momentum parallel to the
projectile axis, P$_{R_{\parallel}}$, is directly related to the
Q-value.

\section{Results:}
A rubidium magneto optical trap (MOT) was used as a target for the charge exchange collisions with cesium projectiles. When the trapping lasers are on, there are two target states available for charge transfer: the ground state (5s) and the first excited state (5p) which is populated by the trapping laser. When the trapping lasers are off, the only available state is the ground state (5s).  Individual cross sections for each charge transfer channel originating from both the rubidium 5s and 5p states were measured, and these cross sections were measured as functions of scattering angle. (At the low collision energies studied in this work, charge capture from the core by singly charged ions is completely negligible.) For these measurements the cesium projectile ion energy was varied from 3.2~keV to 6.4~keV.

\begin{figure}
\centering \fbox{

\begin{minipage}[c]{4.5in}
\includegraphics[width=\textwidth]{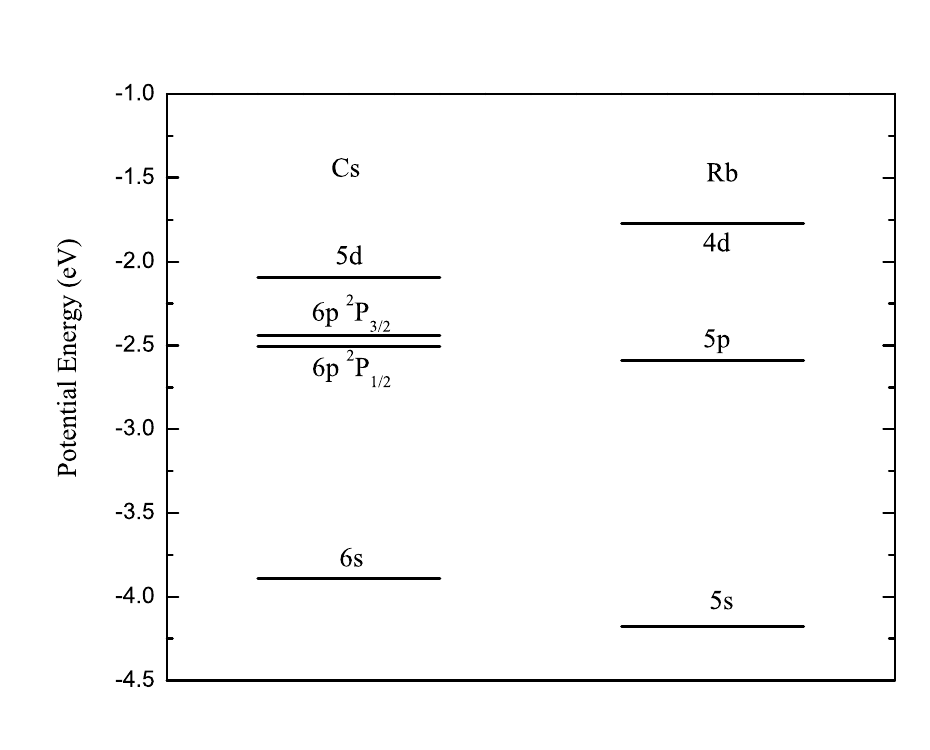}
\end{minipage}}
\caption{Partial energy level diagram for Cesium and Rubidium.}
\label{Fig:CsRbLevels}
\end{figure}

In Fig.~\ref{Fig:CsRbLevels}, the relevant channels for charge exchange
from the trapped rubidium target to the cesium projectile are shown. One
expects that, at low energies, the quasi resonant charge exchange channel
from the ground state is the Rb$(5s)$-Cs$(6s)$ because the electron requires
no additional energy to make the transfer. Similarly, the
quasi-resonant channel for capture from the first excited state is the
Rb$(5p)$-Cs$(6p)$. Charge exchange via other channels is
significantly less, based on this energy pooling argument.
Furthermore, at higher collision energy, charge exchange from
Rb$(5s)$-Cs$(6p)$ should become more pronounced because of the
available collision energy to make the transfer.

Knowledge of excited state fraction is critical in the measurement of relative charge transfer cross sections from excited targets. Because measured charge transfer rates are proportional to the product of the cross section for transfer from some state and the population of that state, one must have an independent measurement of the relative populations in order to obtain the cross sections. Here, the excited state fraction is determined \textit{in situ} for every projectile energy~\cite{Lee2002, Nguyen2004b}. The conditions essential for the success of this method are that the relative capture rates from, and into the various channels can be distinguished, and that the target is cold.  Both conditions are met by MOTRIMS.

By chopping the target excitation (trapping) lasers and comparing the change in charge transfer rates from the ground and excited states for lasers on and lasers off, one can determine the excited state fraction and, independently, the relative capture cross section from both states. If $A_i$ refers to the area under a Q-value peak corresponding to charge transfer from the target's $i^{th}$ initial state whose relative population is given by $n_i$, to a particular final state, then
\begin{equation}
A_s \propto \sigma_s n_s, \label{rate1a}
\end{equation}
\begin{equation}
A_p \propto \sigma_p n_p, \label{rate1b}
\end{equation}
where the constant of proportionality contains acquisition time and geometric factors.  With a high enough chopping frequency, in the case of this work greater than 10~kHz,
\begin{equation}
n_s + n_p = constant ,
\label{const_n}
\end{equation}
because the cold atoms have no time to leave the collision region.  Therefore
\label{allrates2}
\begin{equation}
\Delta A_s \propto \sigma_s \Delta n_s \label{rate2a}
\end{equation}
\begin{equation}
\Delta A_p \propto \sigma_p \Delta n_p  \label{rate2b}
\end{equation}
\begin{equation}
\label{sum_deltas2} \Delta n_s + \Delta n_p = 0,
\end{equation}
where $\Delta n_i$ refers to changes in the $i^{th}$
population as the trapping (and/or re-pumping) laser goes from the
on condition to the off condition.  Note that it is the low
temperature of the target which allows Eqs.~[\ref{const_n}] and
[\ref{sum_deltas2}] to be satisfied at relatively low chopping
frequencies.

Taking the ratio of Eq.~[\ref{rate2b}] to Eq.~[\ref{rate2a}], and
using Eq.~[\ref{sum_deltas2}] we obtain
\begin{equation}
\label{final_sig1} \frac{\Delta A_p}{\Delta A_s} =
\frac{\sigma_p}{\sigma_s}\frac{\Delta n_p}{\Delta
n_s}=-\frac{\sigma_p}{\sigma_s}.
\end{equation}
Then, using this with Eqs.~[\ref{rate1a}],~[\ref{rate1b}], and~[\ref{const_n}],
\begin{equation}
\label{levels1}
\frac{n_p}{n_s}=\frac{\sigma_s}{\sigma_p}\frac{A_p}{A_s}=-\frac{A_p}{A_s}
\frac{\Delta A_s}{\Delta A_p}.
\end{equation}
The method can be generalized to an arbitrary number of excited levels in the target, the only limitation being the Q-value resolution of the MOTRIMS technique.

\begin{figure}
\includegraphics[width=\textwidth]{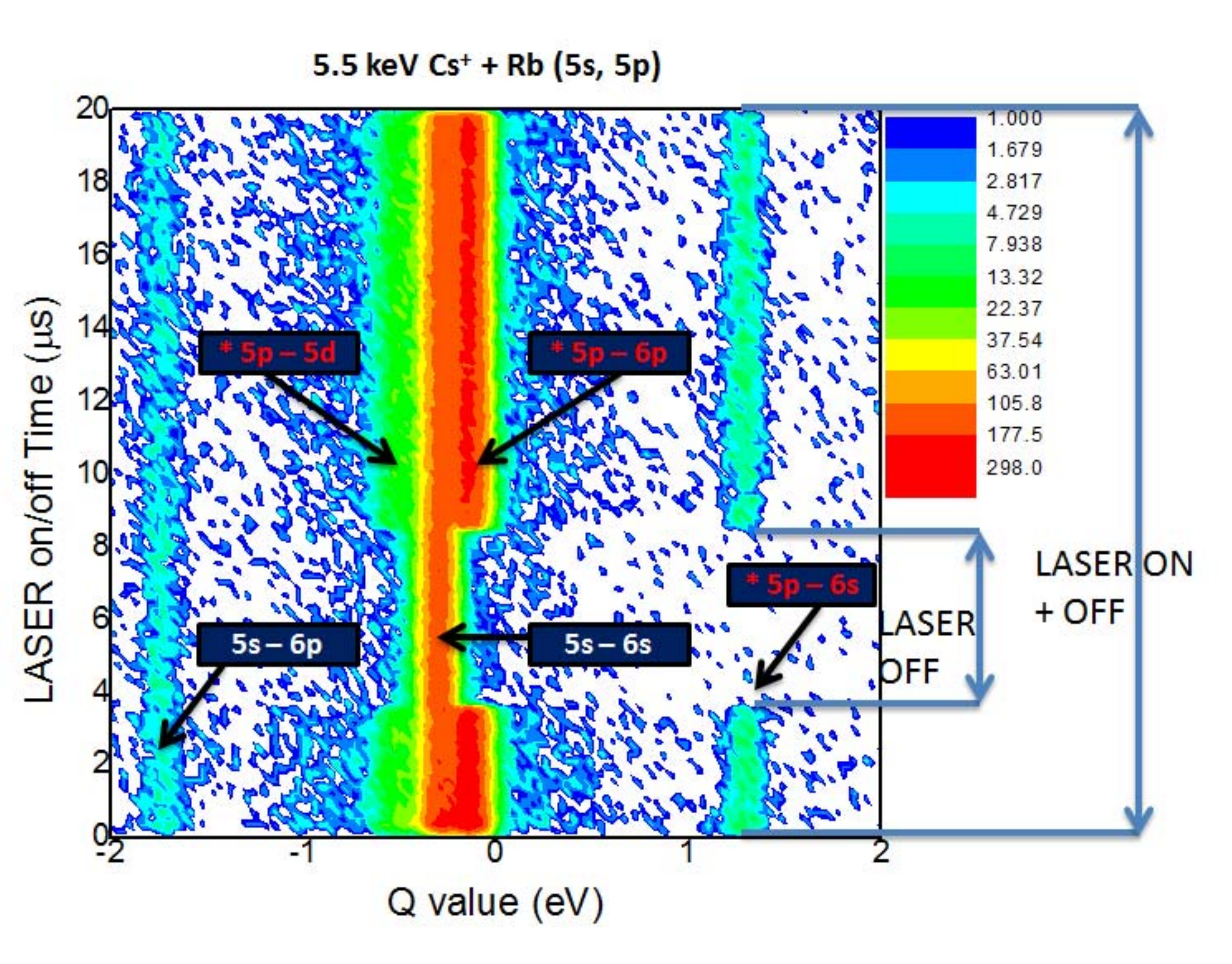}
  \caption{Charge transfer counts are plotted {\it versus} Q-value (eV) and time in the laser on/off cycle ($\mu$s).  ``*'' on the labels indicate charge exchange from Rb(5p). Using data plotted in this manner allows the determination of the excited state fraction in the Rb target.}
  \label{Fig:QvaluesAOM}
\end{figure}

An acousto optical modulator (AOM) was used to quickly chop the trapping lasers on and off. Typically, the lasers are off for 5~$\mu$sec and on for 15~$\mu$sec. A timing pulse, synchronous with the AOM controller, was used as the START for a time-to-amplitude converter (TAC).  A Rb ion detected in coincidence with a neutral projectile provided the STOP signal for the TAC.  Thus, for every projectile energy investigated, we obtained a spectrum of our TAC signal which record the laser on/off time as a function of TOF signal which translates into Q-value via Eqs.~[\ref{Eq:QV}].  A sample spectrum is shown in Fig.~\ref{Fig:QvaluesAOM}.  Vertical integration of this spectrum during laser-on time yields a Q-value spectrum that can be compared to the corresponding Q-value spectrum taken from the laser-off portion of Fig.~\ref{Fig:QvaluesAOM}.  The analysis of Eqs.~[\ref{final_sig1}] and [\ref{levels1}] was then applied to the pairs of laser-on, laser-off Q-value spectra.  Typically, the target excited state fraction was about 20\%.

\begin{figure}
\includegraphics[width=\textwidth]{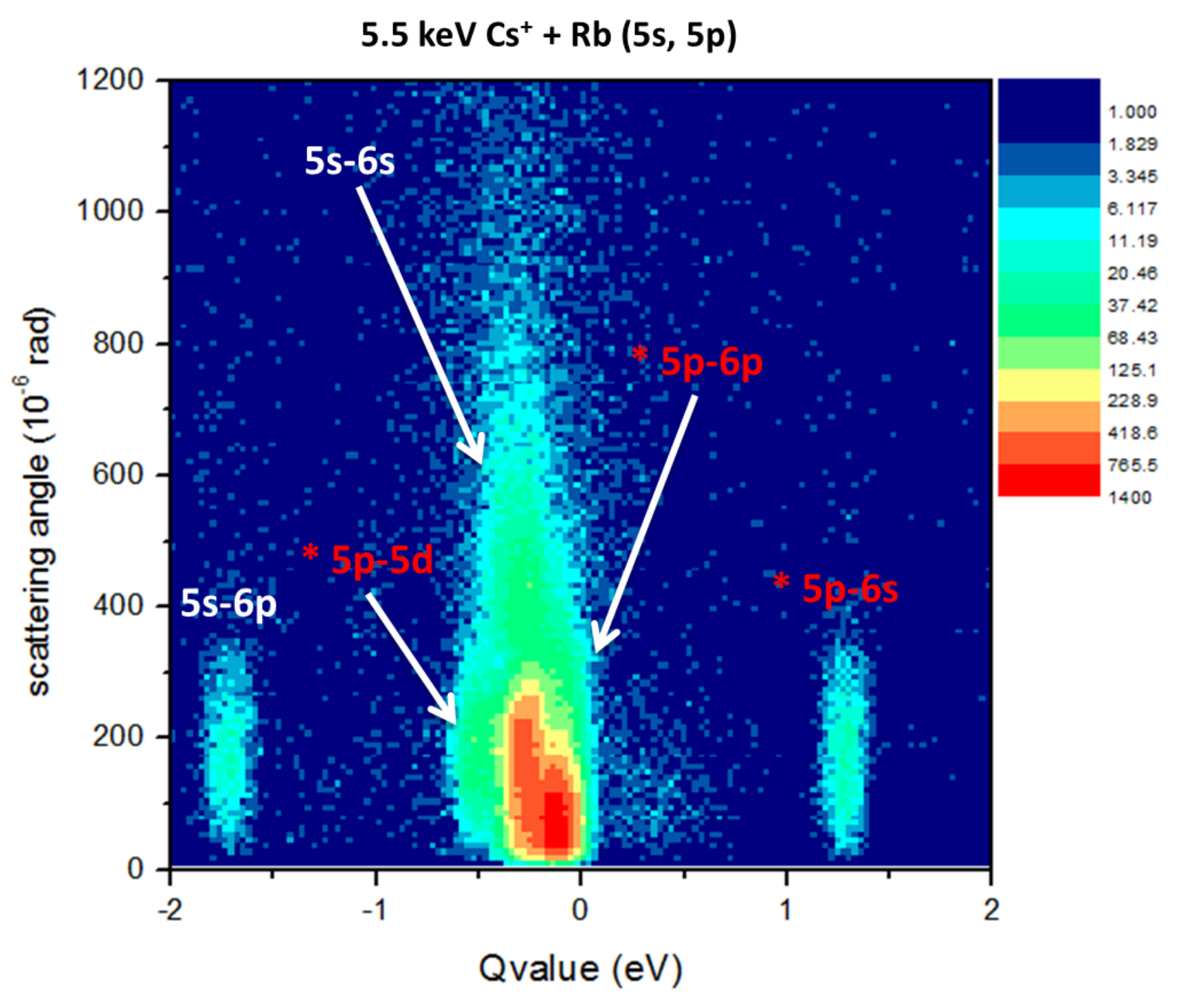}
  \caption{For every projectile energy, cross sections differential in Q-value scattering angles were obtained. Shown here are data for 5.5 keV Cs$^{+}$ + Rb(5s, 5p). ``*'' on the labels indicate charge exchange from Rb(5p).}
  \label{Fig:CsRbDoubleDifferentialCS}
\end{figure}

\begin{figure}
     \includegraphics[width=\textwidth]{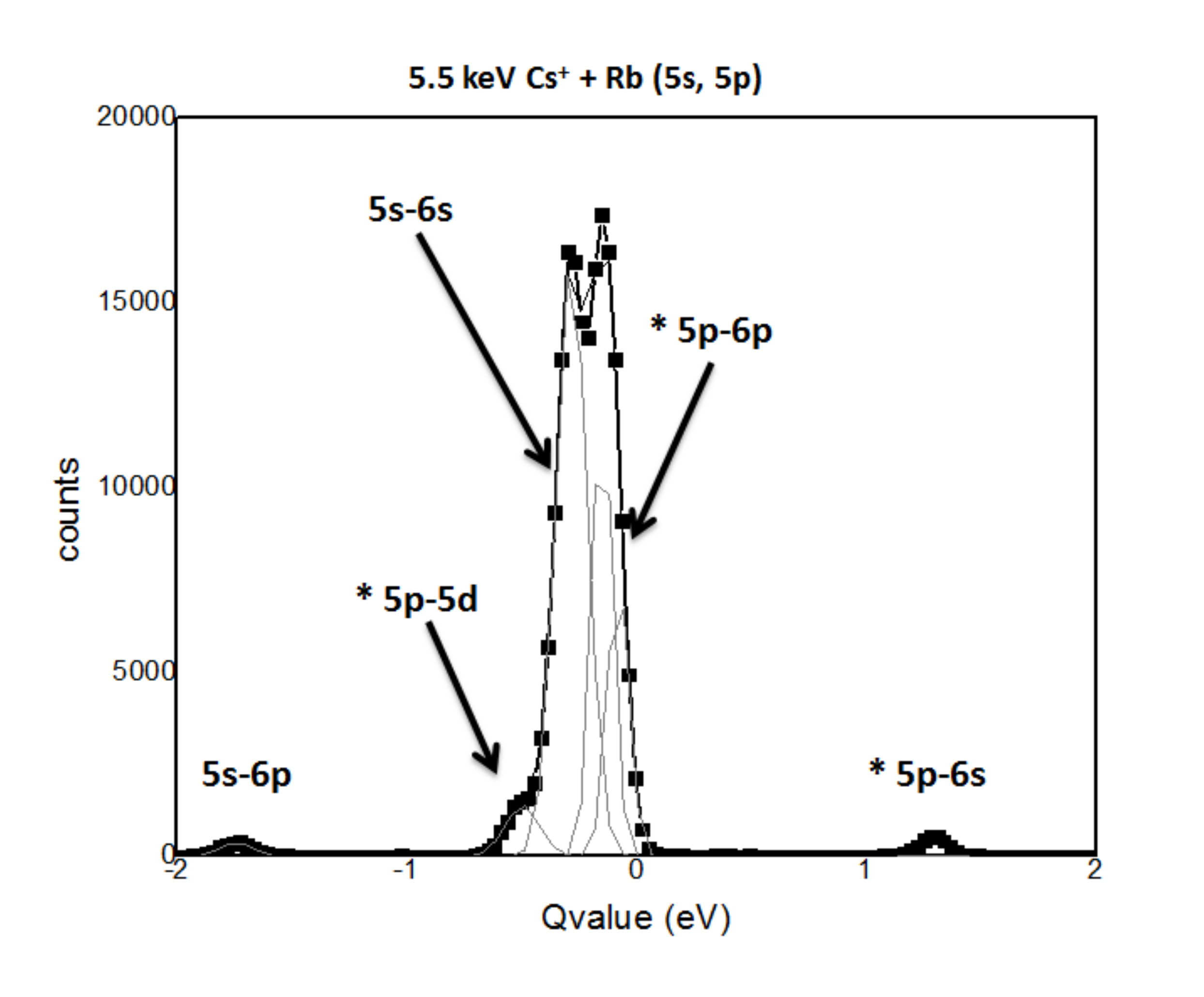}
  \caption{A typical plot of counts {\it versus} Q-value, or energy defect. The labels identify the final state in cesium; ``*'' on the labels indicate charge exchange from Rb(5p).}
\label{Fig:CsRbQValue}
\end{figure}

For every projectile energy investigated, we also obtained a spectrum such as the one shown in Fig.~\ref{Fig:CsRbDoubleDifferentialCS}. This figure, also synchronized to laser-on/laser-off, shows relative cross sections, differential in Q-value and scattering angle, for a typical projectile energy of 5.5~keV. After identifying the channels of interest within a particular range of Q-values, we can make downward projection to obtain the Q-value peaks as shown in Fig.~\ref{Fig:CsRbQValue}. The resolution in Q-value is limited by the detector timing, 2 nanoseconds, which translates into about 100~meV for this collision system.

For the range of energies investigated here, as expected, the
dominant electron capture channel cross sections from ground state
Rb is Rb(5s)-Cs(6s). The remaining contribution to the total capture
from ground state is the Rb(5s)-Cs(6p) channel.  Over the range of
projectile energies studied, the relative capture cross sections
remain constant for both channels.  The relative capture cross
sections, normalized to the total capture from the ground state
Rb(5s) were measured to be $0.97 \pm 0.03$ and $0.03 \pm 0.01$ for
the Rb(5s)-Cs(6s) and Rb(5s)-Cs(6p) channels, respectively.

\begin{figure}
\includegraphics[width=\textwidth]{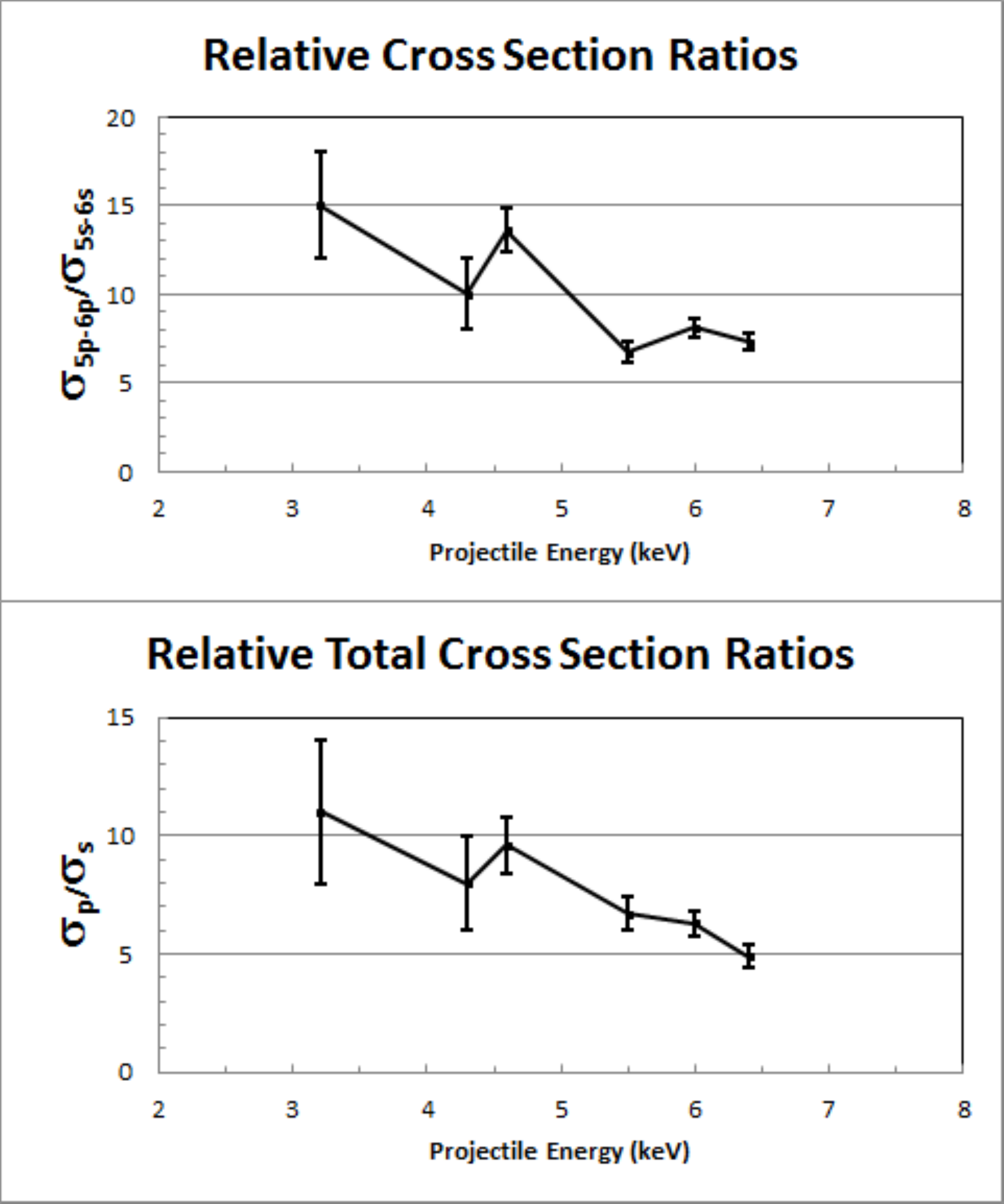}
  \caption{The upper panel shows the experimental energy dependent relative cross section ratios. The lower panel shows the experimental energy dependent relative total cross section ratios.}

  \label{Fig:energydependentrelativecrosssection}
\end{figure}

Table~\ref{Tab:CsRelP} shows the relative capture cross sections
from the first excited Rb(5p) state to various final states. Charge
transfer from Rb(5p) to Cs(6p) is quasi-resonant and therefore
dominates. Table~\ref{Tab:CsRelppss} shows the energy-dependent
relative cross section ratios between the dominant channel for
capture from Rb(5p) to that from Rb(5s) as well as energy-dependent relative total cross section ratios for capture from Rb(5p) to capture from Rb(5s); these data are also plotted
in Fig.~\ref{Fig:energydependentrelativecrosssection}. In general, as projectile energy increases, the total cross section
for capture from Rb(5p) decreases with respect to capture from
Rb(5s).


%
%

\begin{table}
  \centering
  \begin{tabular}{|c|c|c|c|}\hline
 & & & \\

{\bf Energy} (keV) & {\bf 5p-6s} & {\bf 5p-5d} &  {\bf 5p-6p} \\
 & & & \\
\hline
 & & & \\
3.2  & 0.03 $\pm$ 0.01 & 0.06 $\pm$ 0.01 &  0.91 $\pm$ 0.03\\
 & & & \\
4.3  & 0.02 $\pm$ 0.01 & 0.06 $\pm$ 0.01 &  0.92 $\pm$ 0.02\\
 & & & \\
4.6  & 0.03 $\pm$ 0.01 & 0.06 $\pm$ 0.01 &  0.90 $\pm$ 0.02\\
 & & & \\
5.5  & 0.03 $\pm$ 0.01 & 0.10 $\pm$ 0.01 &  0.87 $\pm$ 0.04\\
 & & & \\
6.0  & 0.02 $\pm$ 0.01 & 0.03 $\pm$ 0.02 &  0.95 $\pm$ 0.07\\
 & & & \\
6.4  & 0.03 $\pm$ 0.01 & 0.07 $\pm$ 0.01 &  0.90 $\pm$ 0.02\\
 & & & \\
\hline

  \end{tabular}
  \caption{Experimental state selective energy dependent relative cross sections
  for the Cs$^{+}$ + Rb(5p) collision system. Here the relative cross
sections for each channel are normalized to the excited state
Rb(5p) total cross section. Note that electron capture 5p-7s channel is too small to be measured with any meaningful certainty.}
\label{Tab:CsRelP}
\end{table}

\begin{table}
  \centering
  \begin{tabular}{|c|c|c|}\hline
        &   &\\
   {\bf Energy} (keV) & {\LARGE $\frac{\sigma_{5p-6p}}{\sigma_{5s-6s}}$ } & {\LARGE $\frac{\sigma_{p}}{\sigma_{s}}$ } \\
        &   &\\
    \hline
        &   &\\
    3.2 & 15  $\pm$ 3 & 11 $\pm$ 3\\
        &   &\\
    4.3 & 10  $\pm$ 2 & 8 $\pm$ 2\\
        &   &\\
    4.6 & 13.6  $\pm$ 1.2 & 9.6 $\pm$ 1.2\\
        &   &\\
    5.5 & 6.7  $\pm$ 0.6 & 6.7 $\pm$ 0.7\\
        &   &\\
    6.0 & 8.1  $\pm$ 0.5 & 6.3 $\pm$ 0.5\\
        &   &\\
    6.4 & 7.3  $\pm$ 0.5 & 4.9 $\pm$ 0.5\\
        &   &\\
    \hline
  \end{tabular}
  \caption{Energy-dependent relative cross section ratios for the dominant channels and relative total cross section ratios for Cs$^{+}$ + Rb(5$l$) where $l$ = s and p.}
  \label{Tab:CsRelppss}
\end{table}

\begin{figure}
\includegraphics[width=\textwidth]{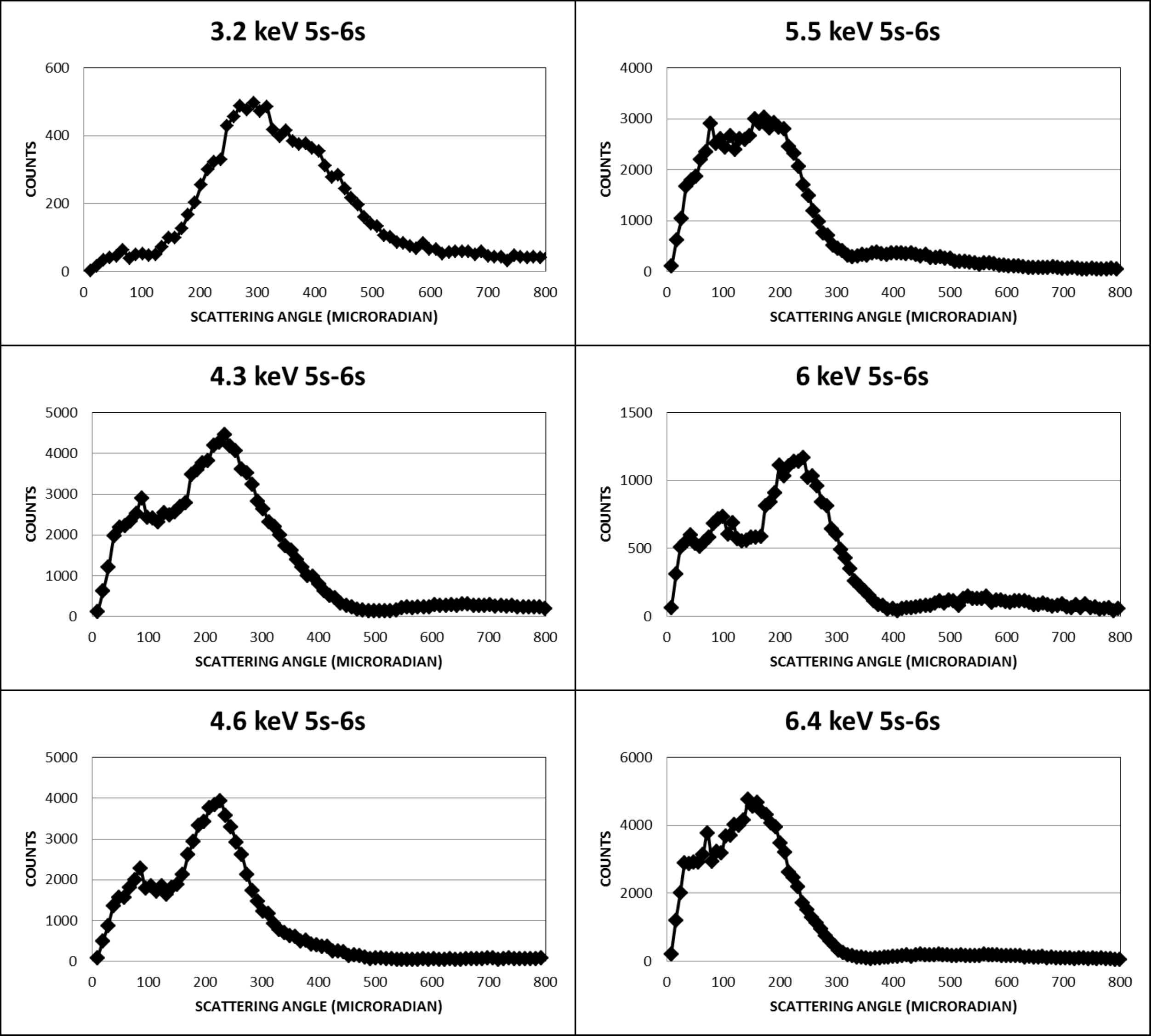}
  \caption{Energy dependent cross sections (arbitrary units) {\it versus} scattering angle for electron charge transfer from Rb (5s) to Cs(6s). The resolution is
approximately 30~microradians. The vertical error is
statistical and they are the size of the black diamond.}
\label{Fig:energydependentscattss}
\end{figure}

\begin{figure}
\includegraphics[width=\textwidth]{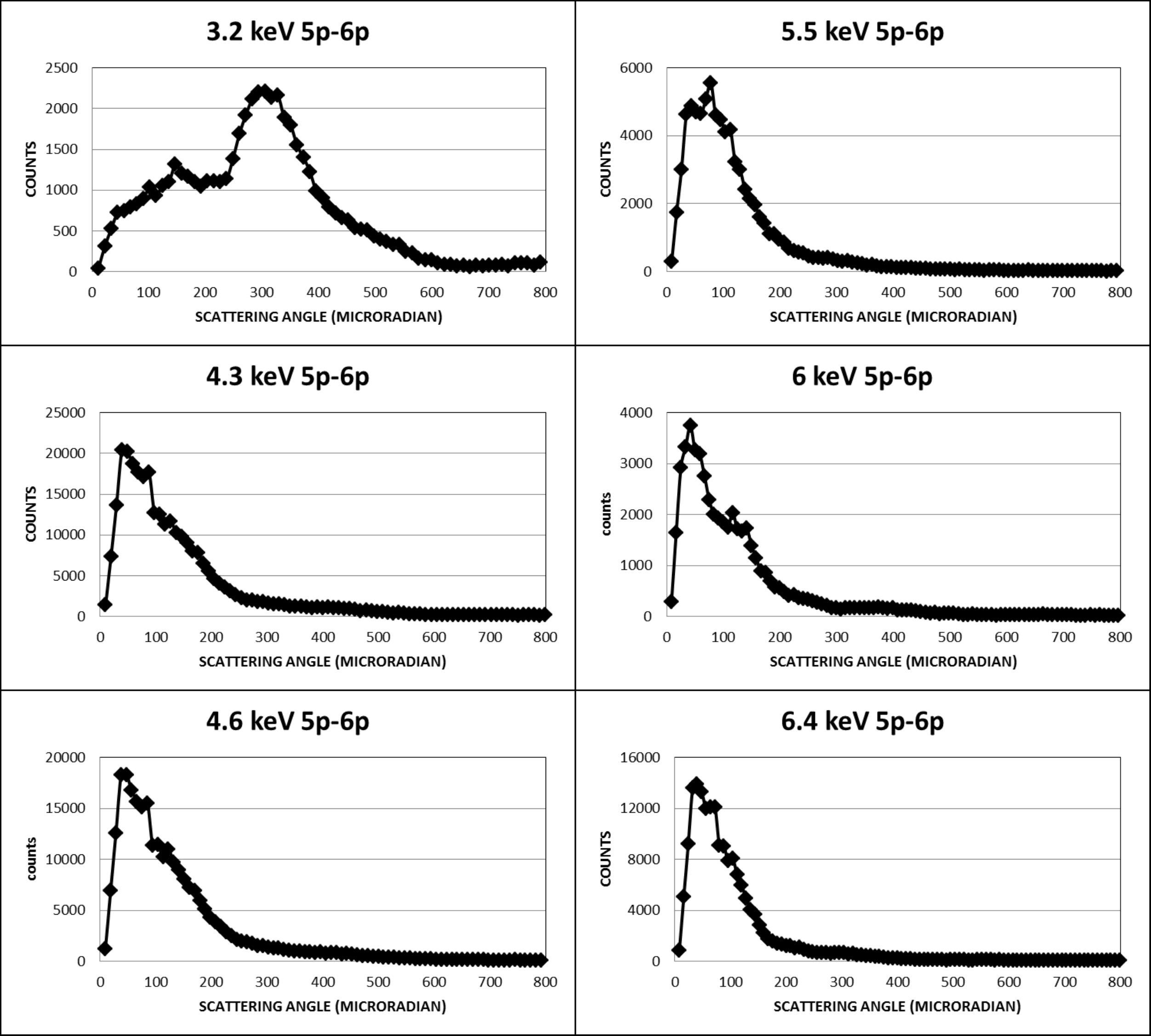}
  \caption{Energy dependent cross sections (arbitrary units) {\it versus} scattering angle for electron charge transfer from Rb (5p) to Cs(6p). The resolution is
approximately 30~microradians. The vertical error is
statistical and they are the size of the black diamond.}\label{Fig:energydependentscattpp}
\end{figure}

Referring again to Fig.~\ref{Fig:CsRbDoubleDifferentialCS}, by selectively gating a small range of Q-values centered on individual capture channels, and projecting horizontally, we obtain the projectile scattering angle information for the channels of interest. Shown in Figs.~\ref{Fig:energydependentscattss} and~\ref{Fig:energydependentscattpp} are energy dependent scattering angles for the dominant channels for capture from Rb(5s) and Rb(5p), respectively.  The resolution in scattering angle, essentially limited by the resolution of the position sensitive detector, is about 30~microradians.  There is some distortion to the PSD image due to the magnetic field gradient of the MOT which, consistent with ion trajectory calculations, leads to a rotation and ``shearing'' of the recoil ion image. This distortion has been corrected in software and does not appear to degrade the resolution in scattering angle.  Energy dependent diffraction-like oscillations reminiscent of data from a different system ~\cite{Leredde2012} can be compared to rigorous two center atomic orbital and molecular close coupling calculations~\cite{Leredde2012, Lee2002}.

\section{Conclusions:}
MOTRIMS was used to make energy-dependent kinematically complete experiments of charge exchange cross sections for the major channels at various projectile energies between Cs$^{+}$ + $^{87}$Rb(5s,5p) in order to provide the most stringent test of theory in the binary collisions with experimental results which are differential in as many parameters as possible. Details of the experimental techniques and data for energy dependent cross sections, differential in Q-value and scattering angle, were given for the two dominant channels, Rb(5s)-Cs(6s)and Rb(5p)-Cs(6p). These high resolution energy dependent measurements, differential in both in Q-value and scattering angle, provide benchmark data which can be used to further develop theoretical treatment. In the projectile energy range investigated here, the other channels remain weakly populated, with no significant dependence on the projectile energy. In contrast, the  ratio between total electron capture cross section from Rb(5p)  and from Rb(5s), was found to decrease rapidly with the collision energy. We further found that the transverse momentum distribution (scattering angle information) showed an oscillatory structure which is sensitive to the projectile energy. Similar to the analysis suggested by Otranto \textit{et. al.}~\cite{Otranto2012}, this feature could be a consequence of the number of swaps the electron undergoes across the potential energy saddle during the charge exchange process.

\section*{Acknowledgments:}
H.~Nguyen would like to thank the University of Mary Washington Research Initiatives and the Summer Science Institute.

\end{document}